# Fractional Fokker-Planck Equation for Ultraslow Kinetics

A. V. Chechkin[1*], J. Klafter[2†], and I. M. Sokolov[3‡]

[1]*Institute for Theoretical Physics, N SC KIPT,
Akademicheskaya st.1, 61108 Kharkov, Ukraine*
[2]*School of Chemistry, Sackler Faculty of Exact Sciences,
Tel Aviv University, Tel Aviv 69978, Israel,
Institute for Theoretical Physics, N SC KIPT,
Akademicheskaya st.1, 61108 Kharkov, Ukraine*
[3]*Institut für Physik, Humboldt-Universität zu Berlin, Invalidenstrasse 110,
D-10115 Berlin, Germany*

**Abstract**

Several classes of physical systems exhibit ultraslow diffusion for which the mean squared displacement at long times grows as a power of the logarithm of time ("strong anomaly") and share the interesting property that the probability distribution of particle's position at long times is a double-sided exponential. We show that such behaviors can be adequately described by a distributed-order fractional Fokker-Planck equations with a power-law weighting-function. We discuss the equations and the properties of their solutions, and connect this description with a scheme based on continuous-time random walks.

    The mean squared displacement of a Brownian particle grows linearly in time, $\langle x^2(t) \rangle \propto t$, a behavior refered to as normal diffusion. In complex systems this kind of behavior is often violated, leading to anomalous diffusion, $\langle x^2(t) \rangle \propto t^\beta$, with $\beta \neq 1$. While the behavior of a diffusing particle under the presence of an external force is well-described by a Fokker-Planck equation (FPE), the corresponding level of description for anomalous diffusion is given by the generalization of a Fokker-Planck scheme based on fractional derivatives and integrals [1-3]. For the subdiffusive case $0 < \beta < 1$ such equations typically include an additional temporal fractional derivative in front of a regular Fokker-Planck operator.

    Other physical situations exist with time evolution which is slower than displayed by a power-law, namely, the mean squared displacement grows as some power of the logarithm of time,

$$\langle x^2(t) \rangle \propto \log^\nu t \ , \tag{1}$$

---

* Electronic address: achechkin@kipt.kharkov.ua

† Electronic address: klafter@post.tau.ac.il

‡ Electronic address: igor.sokolov@physik.hu-berlin.de

with $\nu > 0$. The most known example of such a logarithmic, ultraslow diffusion, also called "strong anomaly" [4], is given by Sinai diffusion for which $\nu = 4$ [5], a case in which the particle moves in a quenched random force field. An even slower behavior with $\nu = 4/3$ is encountered in polymer physics (a polyampholyte hooked around an obstacle [6]). Similar behaviors were observed for motion in aperiodic environments [7] and in a family of iterated maps [4]. Strong anomaly in diffusion was also found numerically in an area preserving parabolic map on a cylinder [8].

Some of the ultraslow systems share the interesting property that the probability density function (PDF) $f(x,t)$ of the displacements scales as a function of $\xi = x/(\log(t))^{\nu/2}$, and displays exponential decay in its wings,

$$f(\xi) \propto \exp(-A|\xi|) \qquad (2)$$

for large $|\xi|$.

A question now arises if one can derive a generalized Fokker-Planck equation which in the limit of no external force leads to Eqs. (1) and (2). In this Letter we demonstrate that the logarithmic laws and the corresponding PDFs can be described within a framework of fractional kinetic equations, but require a new type of approach involving so called distributed-order derivatives, which were first introduced by Caputo [9-11].

Parallel to Ref.[12] we write the distributed order fractional FPE for the PDF $f(x,t)$ as

$$\int_0^1 d\beta \tau^{\beta-1} p(\beta) \frac{\partial^\beta f}{\partial t^\beta} = \hat{L}_{FP} f(x,t) \quad, \quad f(x,0) = \delta(x) \quad, \qquad (3)$$

where $\tau$ is a positive constant representing some characteristic time of the problem (vide infra), $[\tau] = \sec$, and $\hat{L}_{FP}$ is the regular Fokker-Planck operator under an external potential $U(x)$,

$$\hat{L}_{FP} = \frac{\partial}{\partial x} \frac{U'(x)}{m\gamma} + D \frac{\partial^2}{\partial x^2} \quad, \qquad (4)$$

where $m$ is the particle mass, $\gamma$ and $D$ are the friction and the diffusion coefficients, respectively.

Let us discuss the construction in the l.h.s. of the equation in more detail. It is known, that the continuation of the notion of a derivative of the integer order to a non-integer $\beta$ may lead to different operators (fractional derivatives) which may differ with respect to how the initial conditions are handled. Here we understand our time fractional derivative of order $\beta$ in the following sense [13],

$$\frac{\partial^\beta f}{\partial t^\beta} = \frac{1}{\Gamma(1-\beta)} \int_0^t dt'(t-t')^{-\beta} \frac{\partial f}{\partial t'} \quad, \qquad (5)$$

where the sequence of temporal integration and differentiation is reversed with respect to a more common Riemann-Liouville operator. The weight function $p(\beta)$ is a dimensionless non-negative function, which should fulfil $\int_0^1 d\beta p(\beta) = 1$. The explicit form of $p(\beta)$ will be discussed in what follows. As proven in [12], by using the method similar to that presented in [14], the solution of Eq.(3) is a PDF.

Distributed-order FFPEs are a versatile tool for describing slow kinetic processes. As we proceed to show, the ultraslow case corresponds to a class of $p(\beta)$ having a simple power-law form,

$$p(\beta) = \nu \beta^{\nu-1} \qquad . \qquad (6)$$

The normalization condition for $p(\beta)$ on $[0, 1]$ assumes $\nu > 0$. The choice of $p(\beta)$, Eq.(6), captures effects found in different models exhibiting ultra-slow kinetics. We hasten to note that only the behavior of $p(\beta)$ in vicinity of $\beta = 0$ is of significance for the long-time behavior of the overall process. However, we confine ourselves to an explicit form, Eq.(6), and discuss the behavior of the solutions in the whole time-domain.

For $p(\beta) = \delta(\beta - 1)$ in Eq.(3) we get the regular FPE, whose properties has been investigated extensively [15]. Setting $p(\beta) = \delta(\beta - \beta_0)$, $0 < \beta_0 < 1$, we arrive at the FFPE, which have been recently introduced leading to [2]:

(i) The stationary solution is given by the Gibbs-Boltzmann distribution $f_{st}(x) = C \exp(-U(x)/k_B T)$, where $C$ is a normalization constant, $k_B$ is the Boltzmann constant, and $T$ is a temperature;

(ii) The Einstein-Stokes-Smoluchowski relation, $D = k_B T / m\gamma$ is fulfilled;

(iii) The second Einstein relation $\langle x(t) \rangle_F = F \langle x^2(t) \rangle_0 / (2k_B T)$ holds: the first moment in presence of the constant force $F$ is proportional to the second moment in absence of this force. In a spectral representation this relation leads to the Nyquist theorem.

Some of these properties are also shared by the distributed-order equations. The first two properties are fulfilled by virtue of Eq.(3), and are not specific for any particular choice of $p(\beta)$. The third one will be proven for our special case by an explicit calculation.

Let us first examine the behavior of the solutions of the distributed-order equation in the force-free limit ($U(x) = 0$). In the Laplace-Fourier representation the solution

$$\hat{\tilde{f}}(k,s) = \int_{-\infty}^{\infty} dx e^{ikx} \int_0^t dt\, e^{-st} f(x,t) \ , \tag{7}$$

of Eq.(3) for $U = 0$ is explicitly given by

$$\hat{\tilde{f}}(k,s) = \frac{1}{s} \frac{I(s\tau)}{I(s\tau) + k^2 D\tau} \ , \tag{8}$$

where

$$I(s\tau) = \int_0^1 d\beta\, (s\tau)^\beta p(\beta) \ . \tag{9}$$

Since we are interested in small and large time asymptotics, it is sufficient to obtain the expressions for $I(s\tau)$ at large and small $s$:

$$I(s\tau) \approx \begin{cases} \nu s\tau / \ln s\tau \ , & s\tau \gg 1 \ , \\ \Gamma(\nu + 1)/[\ln(1/s\tau)]^\nu \ , & s\tau \ll 1 \ . \end{cases} \tag{10}$$

Here, the large $s$ asymptotics is obtained via the Laplace method [16]. From Eq.(7), by taking the inverse Fourier transform, we get

$$\tilde{f}(x,s) = \frac{1}{\sqrt{4D\tau}} \frac{I^{1/2}(s\tau)}{s} \exp\left\{ -\frac{|x|}{\sqrt{D\tau}} I^{1/2}(s\tau) \right\} \ . \tag{11}$$

Now, using Eqs.(6) and (9), we obtain for the small and large $t$ asymptotics of the PDF,

$$f(x,t) \approx \sqrt{\frac{\nu}{4D}} \frac{1}{(t \ln(t/\tau))^{1/2}} \exp\left\{ -\sqrt{\frac{\nu}{D}} \frac{|x|}{(t \ln(t/\tau))^{1/2}} \right\} \ , \ t/\tau \ll 1 \ , \tag{12}$$

$$f(x,t) \approx \frac{1}{\sqrt{4D\tau}} \left[ \frac{\Gamma(\nu+1)}{\ln^\nu(t/\tau)} \right]^{1/2} \exp\left\{ -\left(\frac{\Gamma(\nu+1)}{D\tau}\right)^{1/2} \frac{|x|}{\ln^{\nu/2}(t/\tau)} \right\} \quad , t/\tau \gg 1 . \qquad (13)$$

Note that our long-time asymptotic behavior shows exactly the double-sided exponential behavior found in models exhibiting ultra-slow kinetics. From Eqs. (11) and (12) the mean squared displacements follow:

$$\left\langle x^2(t) \right\rangle \propto \begin{cases} \dfrac{2D}{\nu} t \ln(\tau/t), & t/\tau \ll 1 \\ \dfrac{2D\tau}{\Gamma(\nu+1)} \ln^\nu(t/\tau), & t/\tau \gg 1 \end{cases} \qquad (14)$$

Thus, strong diffusion anomalies are described within the proposed formalism. Moreover we have shown that the exponent $\nu$ in Eq.(1) is the one in Eq.(6). We also note the universality of the time-dependence for small $t$, apart from a numerical factor. The behavior of the mean square displacement at large $t$ coincides with the result obtained in [4] by using the CTRW approach for the diffusion generated by iterated maps.

Our results here can be contrasted with the known behavior of the solution of the fixed-$\beta$ FFPEs, for which $\left\langle x^2(t) \right\rangle \propto t^\beta$. In the force-free limit the solution reads $f(x,t) = 2^{-1} D_1^{-1/2} t^{-\beta/2} M_{\beta/2}\left(|x|/(D_1 t^\beta)^{1/2}\right)$ [17], where $M_\nu$ denotes the so-called $M$ function (of the Wright type) of order $\nu$ and $D_1 = D\tau^{1-\beta}$. The $M_\nu$-functions exhibit stretched exponential behavior, $M_\nu(r) \approx A(\nu) Y^{\nu-1/2} \exp(-Y)$, $Y = (1-\nu)\left(\nu^\nu r\right)^{1/(1-\nu)}$ which differs considerably from the double-sided simple exponential decay found in strongly anomalous case. Note that the case $\beta = 1$ corresponds to $M_{1/2}(r) = \pi^{-1/2} \exp\left(-r^2/4\right)$ which is the Gaussian.

Many physical situations can be adequately described in the language of decaying modes, corresponding to the eigenfunctions of the Fokker-Planck operator. Returning to Eq.(3) and introducing a separation ansatz:

$$f(x,t) = T(t)\varphi(x) , \qquad (15)$$

we see that Eq.(3) is reduced to

$$\hat{L}_{FP}\varphi_n + \lambda_n \varphi_n = 0 ,$$
$$\int_0^1 d\beta \tau^{\beta-1} p(\beta) \frac{d^\beta T_n}{dt^\beta} + \lambda_n T_n = 0 . \qquad (16)$$

Taking the Laplace transform of Eq.(16) and using Eqs.(6) and (9), we get

$$\tilde{T}_n(s) \approx \begin{cases} \dfrac{T_n(0)}{s}\left[1 - \lambda_n \tau \dfrac{\ln(s\tau)}{\nu s\tau}\right] , & s\tau \gg 1 \\ \dfrac{T_n(0)}{\lambda_n s\tau[\ln(1/s\tau)]^\nu} , & s\tau \ll 1 \end{cases} , \qquad (17)$$

and using Tauberian theorems, we get

$$T_n(t) \approx \begin{cases} T_n(0)\left(1 - \dfrac{\lambda_n}{\nu} t \ln\left(\dfrac{\tau}{t}\right)\right) , & t/\tau \ll 1 \\ \dfrac{T_n(0)}{\lambda_n \tau \ln^\nu(t/\tau)} , & t/\tau \gg 1 \end{cases} . \qquad (18)$$

Thus, the modes show an ultraslow, logarithmic, decay pattern. This can be again contrasted with the behavior found in fixed-$\beta$ equations for subdiffusion, where the relaxation of single modes is governed by a Mittag-Leffler pattern, $T_n(t) = T_n(0)E_\beta\left(-\lambda_n \tau(t/\tau)^\beta\right)$, which is an exponential decay for $\beta = 1$ and a power-law asymptotic decay, $T_n(t) \propto t^{-\beta}$, for $0 < \beta < 1$.

Let us now consider the response of our system to an external force (bias). While the behavior of a Sinai model in the absence of external bias is adequately described by distributed-order fractional equation, the behavior of the system under such bias is quite different. Thus, the genuine Sinai process under the action of a homogeneous external force $F$ shows the mean displacement that grows as a power of time, $\langle x \rangle \propto t^\mu$ [18]. As we show here, in the case of systems described by the distributed-order fractional equation, this displacement grows considerably slower, namely as a power of *logarithm* of time. Moreover, we are able to prove that our ultraslow kinetics fulfills the second Einstein relation. Consider the first moment in presence of a constant force, $F =$ const. Eq.(3) takes the form

$$\nu \int_0^1 d\beta \tau^{\beta-1} \beta^{\nu-1} \frac{\partial^\beta f}{\partial t^\beta} = -\frac{F}{m\gamma}\frac{\partial f}{\partial x} + D\frac{\partial^2 f}{\partial x^2} \quad . \tag{19}$$

Eq.(19) is the main result here.

We are interested in $\langle x(t) \rangle_F = \int_{-\infty}^\infty dx\, x f(x,t)$. For the distributed order FFPE

$$\langle \tilde{x}(s) \rangle_F = \frac{F\tau}{m\gamma}\frac{1}{sI(s\tau)} \quad , \tag{20}$$

and, after inserting Eq.(9) into Eq.(16) and using Tauberian theorems,

$$\langle x(t) \rangle_F \propto \begin{cases} \dfrac{F}{m\gamma\nu} t \ln\left(\dfrac{\tau}{t}\right) , & t/\tau \ll 1 \\[2mm] \dfrac{F\tau}{m\gamma\Gamma(\nu+1)} \ln^\nu(t/\tau) , & t/\tau \gg 1 \end{cases} . \tag{21}$$

By comparing Eqs.(14) and (21) we get the second Einstein relation:

$$\langle x(t) \rangle_F = \frac{F\langle x^2(t) \rangle_0}{2k_B T} \quad , \tag{22}$$

where $k_B T = m\gamma D$.

The difference between the behavior of the genuine Sinai model and the above result for distributed order FFPE is connected with the fact that in the Sinai model (being a random force model) the external bias strongly perturbs the overall potential in which the particle moves. The model described by the distributed order FFPE is essentially a trap model with a broad distribution of trapping times due to very deep traps. Thus, it should not be a surprise that in our case the perturbation introduced by an external force can be considered as weak, so that the second Einstein's relation holds.

The physical relevance of the fixed-$\beta$ FFPEs is to a large extent due to the fact that they can be viewed as describing a "long-time" limit of continuous time random walks (CTRW), a model which was successfully applied for the description of anomalous diffusion phenomena in many areas, e.g., turbulence [19], disordered media [20], intermittent chaotic systems [21], contaminant dispersion in catchments [22] etc. The CTRW models described by the fixed-$\beta$ equations correspond to the power-law distributions of waiting times between the subsequent steps [2, 23]. Here the power-law behavior $\langle x^2(t) \rangle \propto t^\beta$ of the mean square displacement typically corresponds to the waiting-time distributions $\psi(t) \propto t^{-1-\beta}$ with $0 < \beta < 1$. These distributions posses no integer moments but have fractional moments $\langle t^q \rangle$ up to the order of $q = \beta$. We now turn to the relation of the ultra-slow kinetics and the CTRW processes and show that the distributed-order equations with $p(\beta)$ given by Eq.(6) also describe a kind of CTRW, but with a different class of waiting-time distributions, possessing no moments at all. Let us recall the basic formula of the CTRW in the Fourier-Laplace space [19]:

$$\hat{\tilde{f}}(k,s) = \frac{1-\tilde{\psi}(s)}{s}\frac{1}{1-\hat{\psi}(k,s)} \quad , \qquad (23)$$

where $\tilde{\psi}(s)$ is the Laplace transform of the waiting-time PDF $\psi(t)$, and $\hat{\psi}(k,s)$ is the Fourier-Laplace transform of the joint PDF of jumps and waiting times $\psi(\xi,t)$. In the simplest case one assumes the decoupled form: $\psi(\xi,t) = \lambda(\xi)\psi(t)$. Moreover, the jump length variance is considered to be finite, so that $\hat{\lambda}(k)$, the Fourier transform of $\lambda(\xi)$, is $\hat{\lambda}(k) \approx 1 - D\tau k^2$ to the lowest order in $k$.

First, consider the cases for which at large $t$ the waiting time PDF $\psi(t)$ possesses long power-law tails, such that the mean waiting time diverges, that is, $\psi(t) = \psi(t|\beta) \approx \beta\tau^\beta / t^{1+\beta}$, $0 < \beta < 1$, and, consequently, $\tilde{\psi}(s) \approx 1 - (s\tau)^\beta$ at small $s$. With these assumptions, taking the inverse Fourier-Laplace transform of Eq.(23), we arrive at the FFPE in the force-free limit. Now we consider the case when there is no simple power-law waiting-time PDF but, instead, a weighted mixture of power-law functions,

$$\psi(t) = \int_0^1 d\beta\, p(\beta)\psi(t|\beta) \quad , \qquad (24)$$

where [0,1] is the whole interval for variations of $\beta$. We note that all waiting-time distributions with $\beta \geq 1$ correspond to the behavior described by the first order derivative. Then, for the $\tilde{\psi}(s)$ we have

$$\tilde{\psi}(s) \approx 1 - \int_0^1 d\beta\, (s\tau)^\beta p(\beta) \quad . \qquad (25)$$

Inserting Eq.(25) into Eq.(23) instead of a fixed-$\beta$ $\tilde{\psi}(s)$, we arrive at Eqs.(8) and (9). For $p(\beta)$ in Eq.(6) we have

$$\psi(t) \propto \frac{1}{t[\log(t/\tau)]^{\gamma+1}} \quad . \qquad (26)$$

Thus, our distributed order FFPE, Eq.(3), corresponds to the limit of the CTRW with an extremely broad waiting-time PDFs, such that there are no finite moments [24].

Let us summarize our findings. The distributed-order fractional Fokker-Planck equations provide an instrument for mathematical description of systems displaying ultra-slow kinetics ("strong anomaly"). This behavior is described by Eq.(3) with weighting-functions being a power-law on [0,1], which reproduces the logarithmic growth of the mean squared displacement and the approximately double-sided exponential form of the PDF. The connection between this description and the CTRW-picture with an extremely broad waiting time distribution has been established.

**ACKNOWLEDGEMENTS**

The authors thank R. Gorenflo and F. Mainardi for fruitful discussions. This work is supported by the INTAS Project 00 - 0847. AC acknowledges the hospitality of the School of Chemistry of the Tel-Aviv University. IS is grateful to the Fonds der Chemischen Industrie for partial financial support.